\documentclass{sapm1}
\usepackage{graphicx,dcolumn,bm,color,cite,amsmath,amssymb}
\begin{document}

\title[Anomalous slowing down of individual human activity]{Anomalous slowing
  down of individual human activity due to successive decision-making
  processes}

\author[A. V. Zhukov and S. Fedotov and R. Bouffanais]{Alexander V. Zhukov$^\dagger$,
  Sergei Fedotov$^*$ and Roland Bouffanais$^\dagger$\thanks{$\dagger$
    Singapore University of Technology and Design,  8 Somapah Road, Singapore
    487372; Corresponding Author: R. Bouffanais, email:
    bouffanais@sutd.edu.sg \newline$*$ School of Mathematics, The University
  of Manchester, Manchester M13 9PL, United Kingdom}} \affil{A. V. Zhukov \&
  R. Bouffanais: Singapore University of Technology and Design, Singapore
  487372, Singapore} \affil{S. Fedotov: School of Mathematics, The University
  of Manchester, Manchester M13 9PL, United Kingdom}

\maketitle

\begin{abstract}
  Motivated by a host of empirical evidences revealing the bursty character of
  human dynamics, we develop a model of human activity based on successive
  switching between an hesitation state and a decision-realization state, with
  residency times in the hesitation state distributed according to a
  heavy-tailed Pareto distribution. This model is particularly reminiscent of
  an individual strolling through a randomly distributed human crowd. Using a
  stochastic model based on the concept of anomalous and non-Markovian L\'evy
  walk, we show exactly that successive decision-making processes drastically
  slow down the progression of an individual faced with randomly distributed
  obstacles. Specifically, we prove exactly that the average displacement
  exhibits a sublinear scaling with time that finds its origins in: (i) the
  intrinsically non-Markovian character of human activity, and (ii) the power
  law distribution of hesitation times.
\end{abstract}

\section{Introduction}
%
The influence of individual behaviors on various complex problems of social
activity represents a key target of modern social complexity science.  A
seminal work in this field is due to Barab\'asi~\cite%
{barabasi05}, however, many aspects of individual human dynamics remain far
from being
understood~\cite{song10:_model,song10:_limit_predic_human_mobil}. The key
pioneering achievement of Ref.~\cite{barabasi05} is the fact that generally
human actions are not governed by the Poisson probability law, as it was
conveniently assumed before to quantify various aspects of social human
activity. Instead, the real distribution is heavy tailed and close to the
Pareto
one~\cite{barabasi05,song10:_model,song10:_limit_predic_human_mobil}. Here, we
consider the implications of such power-law distributions of human activity
and demonstrate that successive decision-making processes can drastically slow
down individual movement through a randomly distributed human crowd.

The behavior of an individual in a crowd environment is a part of the more
general problem of crowd behavior as a
whole~\cite{helbing00:_simul,helbing14:_how_save_human_lives_compl_scien}.
Every individual in a crowd will certainly have his own goal and make the
corresponding decisions independently resulting in so-called social
forces~\cite{helbing00:_simul}. It has become an important problem for
different researchers, from psychologists to architects and urban planners, to
be able to predict crowd behaviors as well as its complex dynamics. Mainstream
investigations were so far concentrated on the simulation of the crowd
behavior as a whole complex system revealing collective effects and
self-organizing phenomena ~\cite{thalmann13:_crowd_simul}. However, the human
motion dynamics of a particular individual within a crowd has been relatively
overlooked. Kirchner's field-based model uses a stochastic approach to the
problem of crowd dynamics with a focus on individual
behaviors~\cite{kirchner02:_simul}. Kirchner's model takes inspiration from
the process of chemotaxis developed by some social
microorganisms~\cite{ben-jacob97:_from_ii}, and accounts for two individual
motives: (i) desire to move toward an initially established target, and (ii)
desire to follow the others in a crowd. A similar idea underlies the
active-walker models used for the simulation of trail formation in pedestrian
dynamics~\cite{burstedde01:_simul}. The above models, however, do not account
for true individual decision-making, but rather deal with the common choice
for all crowd members.

The ability of cognizant organisms to make decisions yields fundamental
differences in their individual and group behaviors as compared to
non-cognizant ones. Indeed, using all our sensory modalities, we collect all
available external signals, process them and eventually make a decision based
on a complex integration process, which is far from being clearly
understood~\cite{petit10:_decis}. The aim of this Letter is to show that
successive decision-making processes can drastically slow down individual
movement towards a goal.  The concept of anomalous and non-Markovian L\'evy
walk~\cite{metzler00} is applied to the movement of an individual through a
crowd---this individual being subjected to the necessity of making decisions
at stochastically-distributed points in both time and space. Various diffusion
models have been applied in the past to the study of human behavior involving
decision-making processes~\cite%
{wagenmakers09:_method_ratcl}. The most popular model, applied to a variety of
two-choice reaction time paradigms, has been developed by
Ratcliff~\cite{ratcliff78}, who essentially modified the Wiener diffusion
process and applied it to various types of decision-making related phenomena
\cite{ratcliff98:_model,ratcliff04:_diffus,ratcliff08}\footnote{Those include
  lexical decision, short-term and long-term recognition memory tasks,
  same/different letter-string matching, numerosity judgements,
  visual-scanning tasks, brightness discrimination etc. For a comprehensive
  review of the Ratcliff model: http://star.psy.ohio-state.edu/coglab/}. The
above-mentioned model and its variations only address single-event two-choice
decision making processes. Single-event multi-choice decision making processes
have primarily been studied in the frame of practical problems such as
tourists hesitation in route planning (see
Refs.~\cite{wong09:_touris,smallman10:_proces} and references therein). In
this Letter, we consider the more general case of a multi-event multi-choice
continuous decision-making process as epitomized by the simple process of a
man strolling through a crowded environment.  However, the general formalism
is actually applicable to many processes in which multiple decision-making
events occur during a certain finite time interval---i.e. when ``waiting
time'' (making the decision) alternates with ``jump'' (decision realization).

\section{Stochastic model of movement through a crowd}
Our stochastic model for the individual moving through a crowd is derived from
the L\'evy walk model~\cite{metzler00}, such that an individual can be in two
distinct states: (i) ``straight-motion'' ($s$) state during which he moves
towards a chosen target with constant velocity, say in the positive $x$-axis
direction, and (ii) ``hesitation'' ($h$) state for which he is immobile while
making a decision for further movement. Although pedestrian movements are two
dimensional, here we restrict ourselves to a one-dimensional (1D)
problem---$x$ is the coordinate along a chosen direction towards a given
target---to allow for a complete analytical study. The mechanism of switching
between $s$ and $h $ states cannot be described deterministically, hence a
stochastic $s$--$h$ switching is considered. While in the $s$-state, the
individual moves with constant speed $v$ along the chosen direction during a
random time $T_{s}$ before switching to the $h$-state, in which he stays
immobile for a random time $T_{h}$ (decision time) until another switch to the
$s$-state occurs. The random character of $T_s$ (resp. $T_h$) is fully
characterized by its probability density function (PDF) $\psi _{s}(\tau )$
(resp. $\psi _{h}(\tau )$). The key concept of ``hesitation'' requires further
elaboration and a univocal definition. Here, we adopt the definition by Cho
\textit{et al.}~\cite{cho06:_onlin}, where hesitation is the preliminary
thought process preceding any decision-making process.

Let us now introduce the space-time PDFs for both states $P_{s}(t,x)$ and
$P_{h}(t,x)$; non-Markovian switching processes~\cite{fedotov08:_non_markov}
are considered given the importance of past actions in human dynamics and
mobility~\cite{song10:_model}. We intend to derive the integro-differential
equations for $P_{s}$ and $P_{h}$ considering specific residence time PDFs
$\psi _{s}(\tau )$ and $\psi _{h}(\tau )$. In our non-Markovian framework, the
balance equations for $P_{s}$ and $P_{h}$ are
\begin{equation}
  P_{s}(t,x) =\int_{0}^{t}i_{h}(t-\tau ,x-v\tau )\Psi _{s}(\tau )d\tau
  +P_{s}^{(0)}(x-vt)\Psi _{s}(t),  \label{1}
\end{equation}
\begin{equation}
  P_{h}(t,x) =\int_{0}^{t}i_{s}(t-\tau ,x)\Psi _{h}(\tau )d\tau  +P_{h}^{(0)}(x)\Psi _{h}(\tau ),  \label{2}
\end{equation}
where $P_{s|h}^{(0)}(x)=P_{s|h}(0,x)$ and $\Psi _{s|h}(t)=\int_{t}^{\infty
}\psi _{s|h}(\tau )d\tau $ denote the corresponding survival
probabilities. The balance equations for the switching rates $i_{s}(t,x)$ and
$%
i_{h}(t,x)$, between $s$ and $h$ states respectively, read
\begin{equation}
  i_{s}(t,x) =\int_{0}^{t}\psi _{s}(\tau )i_{h}(t-\tau ,x-v\tau )d\tau
  +P_{s}^{(0)}(x-vt)\psi _{s}(t),  \label{3}
\end{equation}
\begin{equation}
  i_{h}(t,x) =\int_{0}^{t}\psi _{h}(\tau )i_{s}(t-\tau ,x)d\tau  +P_{h}^{(0)}(x)\psi _{h}(t).  \label{4}
\end{equation}
The Master equations for $P_{s}$ and $P_{h}$ are obtained by differentiating
Eqs.~\eqref {1} and \eqref{2} with respect to time:
\begin{align}
  \frac{\partial P_{s}}{\partial t}+v\frac{\partial P_{s}}{\partial x}%
  &=-i_{s}(t,x)+i_{h}(t,x),  \label{5}\\
  \frac{\partial P_{h}}{\partial t}&=-i_{h}(t,x)+i_{s}(t,x).  \label{6}
\end{align}
The switching rates $i_{s}(t,x)$ and $i_{h}(t,x)$ can be found by means of the
Laplace transform
\begin{align}
  i_{s}(t,x)&=\int_{0}^{t}K_{s}(t-u)P_{s}(u,x-v(t-u))du,  \label{7}\\
  i_h(t,x)&=\int_0^t K_h(t-u)P_h(u,x)du, \label{8}
\end{align}
where the memory kernels $K_s(t)$ and $K_h(t)$ have the standard
representations
\begin{equation}
  \tilde{K}_s(s)=\frac{\tilde{\psi}_s(s)}{\tilde{\Psi}_s(s)},\quad \tilde{K}%
  _h(s)=\frac{\tilde{\psi}_h(s)}{\tilde{\Psi}_h(s)},  \label{9}
\end{equation}
the Laplace transform being denoted by the tilde superscript.

\section{Characterization of the subdiffusive displacement}
To quantify the walker's displacement, we seek its average position $\langle
x(t)\rangle $.  In line with empirical observations on human dynamics reported
in Refs.~\cite{barabasi05,song10:_model,barabasi05}, we take the hesitation
time PDF $\psi _{h}(\tau )$ to be a power-law distribution:
\begin{equation}
  \psi _{h}(\tau )\sim \left( \frac{\tau _{h}}{\tau }\right) ^{1+\mu },\quad
  0<\mu <1  \label{10}
\end{equation}%
as $\tau \rightarrow \infty $, $\mu $ being the anomalous exponent and $\tau
_{h}$ is a time scale. The Laplace transform $\tilde{\psi}_{h}(s)$
corresponding to (\ref{10}) can be approximated by
\begin{equation}
  \tilde{\psi}_{h}(s)\sim 1-\left( \tau _{h}s\right) ^{\mu },\quad 0<\mu <1
  \label{11}
\end{equation}%
for small $s$. As expected, the mean hesitating time $\langle T_h \rangle
=\int_{0}^{\infty }\tau \psi _{h}(\tau )d\tau $ is infinite in this case.  For
the $s$ state, since we consider a randomly-distributed crowd, the process of
facing an obstacle or a decision to make can be considered to be
Poissonian. Thus the PDF $\psi _{s}(\tau )$ is assumed to be exponential:
\begin{equation}
  \psi _{s}(\tau )=\nu _{s}e^{-\nu _{s}\tau },  \label{psi_s}
\end{equation}%
with a constant and finite switching rate $\nu _{s}=1/\langle T_s
\rangle$. Its Laplace transform reads
\begin{equation}
  \tilde{\psi}_{s}(s)=\frac{\nu _{s}}{\nu _{s}+s}.  \label{12}
\end{equation}
Let us show that our model predicts a subballistic behavior $\langle
x(t)\rangle \sim t^{\mu }$ with $0<\mu <1$. The Laplace transform of the mean
displacement $%
\langle x(t)\rangle $ is
\begin{equation}
  \langle \tilde{x}(s)\rangle =i\frac{dP(s,k)}{dk}|_{k=0},  \label{13}
\end{equation}%
where $P(s,k)$ is the Laplace--Fourier transform of $%
P(t,x)=P_{s}(t,x)+P_{h}(t,x)$ defined by
\begin{equation}
  P(s,k)=\int_{\mathbb{R}}\int_{0}^{\infty }e^{-ikx+st}P(t,x)dt\,dx.  \label{14}
\end{equation}%
From Eqs. \eqref{5}--\eqref{8}, $P(s,k)$ can be explicitly derived as
\begin{eqnarray}
  P(s,k) &=&P_{s}^{(0)}(k)\frac{\tilde{\Psi}_{s}(s+ikv)+\tilde{\Psi}_{h}(s)%
    \tilde{\psi}_{s}(s+ikv)}{1-\tilde{\psi}_{h}(s)\tilde{\psi}_{s}(s+ikv)}+
  \nonumber \\
  &&\ \ \ \ \ \ P_{h}^{(0)}(k)\frac{\tilde{\Psi}_{h}(s)+\tilde{\psi}_{h}(s)%
    \tilde{\Psi}_{s}(s+ikv)}{1-\tilde{\psi}_{h}(s)\tilde{\psi}_{s}(s+ikv)}.
  \label{15}
\end{eqnarray}%
Using Eqs.~\eqref{11} and \eqref{12} inside Eq.~\eqref{15}, together with
$\tilde{\Psi}_{s|h}(s)=(1-\tilde{\psi}_{s|h}(s))/s$ , we obtain the average
position $\langle x(t)\rangle $ in the limit $t\rightarrow \infty $:
\begin{equation}
  \left\langle x(t)\right\rangle \sim \frac{vt^{\mu }}{\Gamma (1+\mu )\nu
    _{s}\tau _{h}^{\mu }},\qquad 0<\mu <1,  \label{18}
\end{equation}%
where $\Gamma$ denotes the classical Gamma function. Interestingly, the
scaling of $\langle x(t) \rangle $ in~\eqref{18} is sublinear. This is due to
\textit{anomalous switching }~\cite{fedotov08:_non_markov} described by
heavy-tailed hesitation residence time PDF \eqref{10} with infinite mean
residence time. It is important to note that if the PDF $\psi_h$ were
considered to be a short-tailed Poisson distribution, $\psi_h(\tau)=\nu_h
e^{-\nu_h \tau}$, with a finite average hesitation time $\langle T_h \rangle
=1/\nu_h$, then the overall process would be Markovian with a linear ballistic
scaling of the mean position $\langle x(t)\rangle $ on time $t$:
\begin{equation}
  \langle x(t)\rangle =\frac{\nu _{h}}{\nu _{h}+\nu _{s}}vt.  \label{16}
\end{equation}

To gain further insight into the appearance of the nonlinear scaling in time
$t^{\mu }$ in Eq.~\eqref{18}, one can also use the following idea. The average
position $\langle x(t)\rangle $ can alternatively be found as the product of
the average number of jumps $\langle N(t)\rangle $ from $h$-state to
$s$-state, and the distance $v\langle T_s\rangle $ covered in the
$s$-state---$\langle T_s\rangle =1/\nu _{s}$ is the average time spent in the
$s$-state according to the PDF~\eqref{psi_s} for $\psi_s$. Then
\begin{equation}
  \left\langle x(t)\right\rangle =\frac{v\langle N(t)\rangle }{\nu _{s}}.
  \label{182}
\end{equation}%
It is well known from the renewal theory (see,
e.g.,~\cite{cox65:_theor_stoch_proces,feller71}), that the Laplace transform
of $P(n,t)=\Pr (N(t)=n)$ is given by
\begin{equation}
  \tilde{P}(n,s)=\frac{\tilde{\psi}_{h}^{n}(s)(1-\tilde{\psi}_{h}(s))}{s}.
  \label{17}
\end{equation}%
Therefore, the Laplace transform of the average number of jumps from
hesitation state $\langle N(t)\rangle$ is
\begin{equation}
  \langle \tilde{N}(s)\rangle =\sum_{n=0}^{\infty }n\tilde{P}(n,s)=\frac{%
    \tilde{\psi}_{h}(s)}{s(1-\tilde{\psi}_{h}(s))}.  \nonumber
\end{equation}%
It follows from \eqref{11} that $\langle \tilde{N}(s)\rangle \sim \tau
_{h}^{-\mu }s^{-(1+\mu )}$ as $s\rightarrow 0$ and
\begin{equation}
  \langle N(t)\rangle \sim \frac{t^{\mu }}{\Gamma (1+\mu )\tau _{h}^{\mu }}.
  \nonumber
\end{equation}%
This formula together with~\eqref{182} allows us to recover the sublinear
scaling in time of $\langle x(t)\rangle$ previously obtained in~\eqref{18}.

Although we consider a simple model, it enables us to uncover the central fact
that this sublinear scaling in time originates from the presence of memory in
the agent's dynamics. Indeed, we found that if past actions do not affect
decision making, i.e. in the Markovian case, the average displacement given by
Eq.~\eqref{16} is purely ballistic. The importance of memory in human dynamics
has already been highlighted based on a host of empirical evidences, albeit
for processes occurring over much longer spatial and temporal ranges as
compared to the ones in our
study~\cite{song10:_model,song10:_limit_predic_human_mobil,vazquez07:_impac}. For
instance, the individual-mobility model proposed by Song~\textit{et
  al.}~\cite{song10:_model,song10:_limit_predic_human_mobil} to explain some
of these empirical evidences is essentially non-Markovian owing to both the
exploration and preferential return mechanisms and was designed to capture the
long-term spatial and temporal scaling patterns. In comparison, the present
model investigates the short-term scaling in individual mobility that is not
captured in Ref.~\cite{song10:_model,song10:_limit_predic_human_mobil}.

Our 1D individual-stroller model has three parameters, $0<\mu<1$, $\tau_h>0$
and $\nu_s=1/\langle T_s \rangle$, the former two associated with the
hesitation state while the latter fully characterizes the $s$ state. Values
for those parameters could easily be obtained from empirical observations
based on trackings of human strolling along a crowded narrow corridor. Other
testable experiments highly relevant to the present study abound in the field
of sports science, and more specifically with the study of some team sports
from the standpoint of complex dynamical
systems~\cite{frencken08:_team,bourbousson10:_space}. Indeed, with the aim of
improving sports performance, scientists have used video-based and electronic
tracking systems to study space-time coordination dynamics during basketball
and soccer games among others~\cite{frencken08:_team,bourbousson10:_space}. To
the best of our knowledge, the data gathered are mostly analyzed in a compound
way, hence delivering team performance indicators. Apparently, without any
change in methods, individual players kinematics could be analyzed, thereby
providing the PDFs $\psi_s$ and $\psi_h$, along with the associated values for
the three parameters of our model. It is worth adding that data from
basketball games would provide an excellent match with the details of our
model, given that, typically, basketball players switch from periods of
straight forward running ($s$ state) with periods during which they are
immobile, dribbling, and hesitating ($h$ state) before eventually completing a
pass.

The slowed-down dynamics that emerges from successive decision-making
processes should obviously not impugn the role and importance of sensory
modalities in real-life decision-making situations. Interestingly, our model
reveals the counterintuitive fact that making multiple (even seemingly right)
decisions over a long period of time contributes to slowing us down in
reaching our goal. This important fact applies to an individual making his way
through a dense crowd, as well as a basketball player on the court or, more
generally, to any agent whose non-Markovian dynamics alternates between two
states such that the residency time in one of these two states is heavy
tailed.

\section{Conclusion}
%
Returning to our initial focus on human traffic flow, our analytical study
offers unique insights into the origin of the slowed-down dynamics of
individuals moving about a space in the presence of obstacles~\cite%
{brogan03:_realis_human_walkin_paths,yanagisawa09:_introd,steffen10:_method,henry10:_learn_navig_throug_crowd_envir,
  dias12:_turnin_angle_effec_emerg_egres,dias13:_inves}. For instance, in
Ref.~\cite{dias12:_turnin_angle_effec_emerg_egres} the authors conducted a
series of experiments with a group of pedestrians walking through angled
corridors. Specifically, they investigated the dependence of the overall speed
of an `average' pedestrian on the corridor angle. It appears that the speed is
strongly dependent on the angle value and decreases drastically at some
critical value $\theta_c$. The latter varies in a range between $\pi/3$ and $%
\pi/2$ depending on the pedestrian's motivation. Obviously, in the particular
case of an individual walker meandering through a multi-angled corridor the
overall dynamics is slowed down and the speed is highly dependent on the
actual sequence of angles. Although this specific experiment does not exactly
match our model---one-dimensional curvilinear path with angles analogous to
randomly distributed individuals in a given human crowd---the dynamics at play
can readily be understood in our framework. Indeed, in our case, the
individual's straightforward motion is disrupted by the presence of fellow
crowd members leading to sequences of hesitation periods of varying durations
distributed according to a power law. In the experimental framework of
Ref.~\cite{dias12:_turnin_angle_effec_emerg_egres}, the individual is
compelled to execute successive different decision-making processes owing to
the different values of the angles imposed. This leads to successive different
decision realizations, which is essentially equivalent to different hesitation
periods followed by decision realizations as per our model.

\section*{Acknowledgments}
Present work was financially supported by the SUTD-MIT International Design
Centre (IDC). SF gratefully acknowledges the support of the EPSRC under Grant
No. EP/J019526/1.



\begin{thebibliography}{10}
\providecommand{\url}[1]{\texttt{#1}}
\providecommand{\urlprefix}{URL }
\expandafter\ifx\csname urlstyle\endcsname\relax
  \providecommand{\doi}[1]{doi:\discretionary{}{}{}#1}\else
  \providecommand{\doi}{doi:\discretionary{}{}{}\begingroup
  \urlstyle{rm}\Url}\fi
\providecommand{\eprint}[2][]{\url{#2}}

\bibitem{barabasi05}
\textsc{A.-L. Barab{\'a}si}, The origin of bursts and heavy tails in human
  dynamics, \emph{Nature} 435:207--211 (2005).

\bibitem{song10:_model}
\textsc{C.~Song}, \textsc{T.~Koren}, \textsc{P.~Wang}, and \textsc{A.-L.
  Barab\'asi}, Modelling the scaling properties of human mobility, \emph{Nature
  Phys.} 6:818--823 (2010).

\bibitem{song10:_limit_predic_human_mobil}
\textsc{C.~Song}, \textsc{Z.~Qu}, \textsc{N.~Blumm}, and \textsc{A.-L.
  Barab\'asi}, Limits of predictability in human mobility, \emph{Science}
  327:1018--1021 (2010).

\bibitem{helbing00:_simul}
\textsc{D.~Helbing}, \textsc{I.~Farkas}, and \textsc{T.~Vicsek}, Simulating
  dynamical features of escape panic, \emph{Nature} 407:487--490 (2000).

\bibitem{helbing14:_how_save_human_lives_compl_scien}
\textsc{D.~Helbing}, \textsc{D.~Brockmann}, \textsc{T.~Chadefaux},
  \textsc{K.~Donnay}, \textsc{U.~Blanke}, \textsc{O.~Woolley-Meza},
  \textsc{M.~Moussa\"{\i}d}, \textsc{A.~Johansson}, \textsc{J.~Krause},
  \textsc{S.~Schutte}, and \textsc{M.~Perc}, How to save human lives with
  complexity science (2014), ArXiv:1402.7011.

\bibitem{thalmann13:_crowd_simul}
\textsc{D.~Thalmann} and \textsc{S.~R. Musse}, \emph{Crowd Simulation}, second
  ed., Springer, 2013.

\bibitem{kirchner02:_simul}
\textsc{A.~Kirchner} and \textsc{A.~Schadschneider}, Simulation of evacuation
  processes using a bionics-inspired cellular automaton model for pedestrian
  dynamics, \emph{Physica A} 312:260--276 (2002).

\bibitem{ben-jacob97:_from_ii}
\textsc{E.~Ben-Jacob}, From snowflake formation to growth of bacterial colonies
  {II}: {C}ooperative formation of complex colonial patterns, \emph{Comtemp.
  Phys.} 38:205--241 (1997).

\bibitem{burstedde01:_simul}
\textsc{C.~Burstedde}, \textsc{K.~Klauck}, \textsc{A.~Schadschneider}, and
  \textsc{J.~Zittartz}, Simulation of pedestrian dynamics using a
  two-dimensional cellular automaton, \emph{Physica A} 295:507--525 (2001).

\bibitem{petit10:_decis}
\textsc{O.~Petit} and \textsc{R.~Bon}, Decision-making processes: The case of
  collective movements, \emph{Behav. Process.} 84:635--647 (2010).

\bibitem{metzler00}
\textsc{R.~Metzler} and \textsc{J.~Klafter}, The random walk's guide to
  anomalous diffusion: a fractional dynamics approach, \emph{Phys. Rep.}
  339:1--77 (2000).

\bibitem{wagenmakers09:_method_ratcl}
\textsc{E.-J. Wagenmakers}, Methodological and empirical developments for the
  {R}atcliff diffusion model of response times and accuracy, \emph{Eur. J.
  Cogn. Psych.} 21:641--671 (2009).

\bibitem{ratcliff78}
\textsc{R.~Ratcliff}, A theory of memory retrieval, \emph{Psych. Rev.}
  85:59--108 (1978).

\bibitem{ratcliff98:_model}
\textsc{R.~Ratcliff} and \textsc{J.~N. Rouder}, Modeling response times for
  two-choice decisions, \emph{Psych. Sci.} 9:347--356 (1998).

\bibitem{ratcliff04:_diffus}
\textsc{R.~Ratcliff}, \textsc{P.~Gomez}, and \textsc{G.~Mc{K}oon}, Diffusion
  model account of lexical decision, \emph{Psych. Rev.} 111:159--182 (2004).

\bibitem{ratcliff08}
\textsc{R.~Ratcliff} and \textsc{G.~Mc{K}oon}, The diffusion decision model:
  Theory and data for two-choice decision tasks, \emph{Neural Computation}
  20:873--922 (208).

\bibitem{wong09:_touris}
\textsc{J.~Y. Wong} and \textsc{C.~Yeh}, Tourist hesitation in destination
  decision making, \emph{Ann. Tourism Res.} 36:6--23 (2009).

\bibitem{smallman10:_proces}
\textsc{C.~Smallman} and \textsc{K.~Moore}, Process studies of tourists’
  decision-making: {T}he riches beyond variance studies, \emph{Ann. Tourism
  Res.} 37:397--422 (2010).

\bibitem{cho06:_onlin}
\textsc{C.-H. Cho}, \textsc{J.~Kang}, and \textsc{H.~Cheon}, Online shopping
  hesitation, \emph{Cyberpsychol. Behav.} 9:261--274 (2006).

\bibitem{fedotov08:_non_markov}
\textsc{S.~Fedotov} and \textsc{V.~M\'endez}, Non-{M}arkovian model for
  transport and reactions of particles in spiny dendrites, \emph{Phys. Rev.
  Lett.} 101:218102 (2008).

\bibitem{cox65:_theor_stoch_proces}
\textsc{D.~R. Cox} and \textsc{H.~D. Miller}, \emph{The Theory of Stochastic
  Processes}, Methuen, London U.K., 1965.

\bibitem{feller71}
\textsc{W.~Feller}, An introduction to probability theory and its applications,
  vol.~2, Wiley, NY, 1971.

\bibitem{vazquez07:_impac}
\textsc{A.~Vazquez}, Impact of memory on human dynamics, \emph{Physica A}
  373:747--752 (2007).

\bibitem{frencken08:_team}
\textsc{W.~G.~P. Frencken} and \textsc{K.~A. P.~M. Lemmink}, Team kinematics of
  small-sided soccer games: A systematic approach, Science and football {VI}:
  Proceedings of the 6th World Congress on Science and Football,, (Editors)
  \textsc{T.~Reilly} and \textsc{F.~Korkusuz} Routeledge, London, 2008  pp.
  161--166.

\bibitem{bourbousson10:_space}
\textsc{J.~Bourbousson}, \textsc{C.~S\`eve}, and \textsc{T.~McGarry},
  Space-time coordination dynamics in basketball: Part 1. intra- and
  inter-couplings among player dyads, \emph{J. Sports Sci.} 28:339--347 (2010).

\bibitem{brogan03:_realis_human_walkin_paths}
\textsc{D.~C. Brogan} and \textsc{N.~L. Johnson}, Realistic human walking
  paths, Proceedings of the 16th International Conference on Computer Animation
  and Social Agents, IEEE Computer Society, 2003.

\bibitem{yanagisawa09:_introd}
\textsc{D.~Yanagisawa}, \textsc{A.~Kimura}, \textsc{A.~Tomoeda},
  \textsc{R.~Nishi}, \textsc{Y.~Suma}, \textsc{K.~Ohtsuka}, and
  \textsc{K.~Nishinari}, Introduction of frictional and turning function for
  pedestrian outflow with an obstacle, \emph{Phys. Rev. E} 80:036110 (2009).

\bibitem{steffen10:_method}
\textsc{B.~Steffen} and \textsc{A.~Seyfried}, Methods for measuring pedestrian
  density, flow, speed and direction with minimal scatter, \emph{Physica A}
  (2010), 1902--1910.

\bibitem{henry10:_learn_navig_throug_crowd_envir}
\textsc{P.~Henry}, \textsc{C.~Vollmer}, \textsc{B.~Ferris}, and
  \textsc{D.~Fox}, Learning to navigate through crowded environments, 2010
  International Conference on Robotics and Automation ICRA, IEEE Robotics and
  Automation Society, 2010  pp. 981--986.

\bibitem{dias12:_turnin_angle_effec_emerg_egres}
\textsc{C.~Dias}, \textsc{M.~Sarvi}, \textsc{N.~Shiwakoti}, and
  \textsc{M.~Burd}, Turning angle effect on emergency egress experimental
  evidence and pedestrian crowd simulation, \emph{Transport. Res. Rec.}
  2312:120--127 (2012).

\bibitem{dias13:_inves}
\textsc{C.~Dias}, \textsc{M.~Sarvi}, \textsc{N.~Shiwakoti},
  \textsc{O.~Ejtemai}, and \textsc{M.~Burd}, Investigating collective escape
  behaviours in complex situations, \emph{Safety Sci.} 60:87--94 (2013).

\end{thebibliography}
\end{document}